\documentclass[twocolumn,superscriptaddress,showpacs,floatfix,aps,prb]{revtex4}
\usepackage{graphicx}
\usepackage{dcolumn}
\usepackage{amsmath}
\usepackage{amssymb}
\usepackage{amsbsy}
\usepackage{textcomp}
\usepackage{units}
\usepackage{color}
\usepackage{units}
\usepackage{ulem}

\begin{document}

\title{Spin-dependent transport in a multiferroic tunnel junction:\\ Theory for Co/PbTiO$_{3}$/Co}

\author{Vladislav S. Borisov}
\affiliation{Institute of Physics, Martin Luther University Halle-Wittenberg, 06099 Halle, Germany}
\affiliation{Max Planck Institute of Microstructure Physics, 06120 Halle, Germany}
\email[Corresponding author:\ ]{vladislav.s.borisov@gmail.com}

\author{Sergey Ostanin}
\affiliation{Max Planck Institute of Microstructure Physics, 06120 Halle, Germany}

\author{Steven Achilles}
\affiliation{Institute of Physics, Martin Luther University Halle-Wittenberg, 06099 Halle, Germany}

\author{J\"urgen Henk}
\affiliation{Institute of Physics, Martin Luther University Halle-Wittenberg, 06099 Halle, Germany}

\author{Ingrid Mertig}
\affiliation{Institute of Physics, Martin Luther University Halle-Wittenberg, 06099 Halle, Germany}
\affiliation{Max Planck Institute of Microstructure Physics, 06120 Halle, Germany}

\date{\today}

\begin{abstract}
Spin-dependent electronic transport through multiferroic Co/PbTiO$_{3}$/Co tunnel junctions is studied theoretically. Conductances calculated within the Landauer-B\"uttiker formalism yield both a large tunnel magnetoresistance (TMR) and a large tunnel electroresistance  (TER). On top of this, we establish a four-conductance state. The conductances depend crucially on the details of the electronic structure at the interfaces. In particular, the spin polarization of the tunneling electronic states is affected by the hybridization of orbitals and the associated charge transfer at both interfaces. Digital doping of the PbTiO$_{3}$ barrier with Zr impurities at the TiO$_{2}$/Co$_{2}$ interface significantly enhances the TMR\@. In addition, it removes the metalization of the barrier.
 
\end{abstract}


\maketitle

\section{Introduction}
Recently, the functionality of magnetic tunnel junctions has been enhanced by replacing the insulating barrier, typically a band insulator such as MgO (e.¸\,g.\ Ref.~\onlinecite{Yuasa04b}), by a ferroelectric. This combination of ferromagnetism in the electrodes and ferroelectricity in the barrier leads to a multiferroic tunnel junction (MFTJ) in which tunnel magnetoresistance (TMR) and tunnel electroresistance (TER) show up simultaneously. By independently switching  the mutual orientation of the magnetizations in the electrodes---say, from parallel (P, $\uparrow\uparrow$) to antiparallel (AP, $\downarrow\uparrow$)---and reversing the ferroelectric polarization in the barrier---say, from left ($\leftarrow$) to right ($\rightarrow$)---the tunnel conductance may take four different values, leading to a nonvolatile four-state memory device \cite{Velev2009,Garcia2010}.

A four-conductance state has been reported recently for Co/PbTiO$_{3}$/LaSrMnO$_{3}$ tunnel junctions \cite{Pantel2012}, in which a perovskite barrier of $\unit[3.2]{nm}$ PbZr$_{0.2}$Ti$_{0.8}$O$_{3}$ (PZT) has been epitaxially grown on the ferromagnetic and almost half-metallic perovskite La(Sr)MnO$_{3}$(001) (LSMO). An intriguing observation is a sign change of the TMR: switching the ferroelectric polarization in PZT from pointing towards LSMO to pointing towards Co reverses the TMR from $\unit[+4]{\%}$ to $\unit[-3]{\%}$.

For magnetic tunnel junctions, e.\,g.\ Fe/MgO/Fe, it has been shown that details of the interfaces play a decisive role for the TMR \cite{Belashchenko05,Heiliger05,Heiliger06a}: their geometry and their electronic structure govern the spin-dependent transport \cite{Bose08}. Concerning Co/PZT interfaces, \textit{ab initio} calculations \cite{Borisov2014} suggest that (i) the interface is Ti(Zr)O$_{2}$-terminated. More precisely, the  interfacial Co atoms are placed in line with O\@. (ii) Reversal of the ferroelectric polarization of PZT may also reverse the spin polarization at the interface. It is conceivable that this switching is responsible for the experimentally observed sign change of the TMR\@. The interface between LSMO and PZT could be PbO/MnO$_{2}$ since the ferroelectric films grow epitaxially and in complete unit cells on clean substrates \cite{Meyerheim2011,Meyerheim2013}; this scenario is supported by first-principles calculations \cite{Borisov2015a}.

To explain the origin of the inverted TMR\footnote{A TMR is called inverted if the conductance for the parallel magnetic configuration of the electrodes is smaller than that for the antiparallel configuration; as a result, the TMR ratio is negative.} in Co/PZT/LSMO, Pantel \textit{et al.}  point out the role of LSMO (Ref.~\onlinecite{Pantel2012}). A spin polarization of the LSMO surface as large as $\unit[95]{\%}$ has been deduced from transport measurements \cite{Bowen2003}. Thus, both interfaces of Co/PZT/LSMO may contribute significantly to the above-mentioned peculiarities of the TMR\@. In Reference~\onlinecite{Pantel2012} a contribution from resonant tunneling\cite{Yuasa02,Wunnicke02b,Tsymbal2003} via electronic states localized within the barrier has been ruled out. We recall that resonant states can alter the spin polarization of the tunneling electrons \cite{Tsymbal2003}.

In this work, we investigate spin-dependent tunneling in Co/PbTiO$_{3}$/Co and Co/PbTiO$_{3}$-Ti(Zr)O$_{2}$/Co MFTJs by means of first-principles electronic structure and transport calculations\cite{Lukashev2013,Xiaohui2013}. These tunnel junctions, with their two Co electrodes, have been chosen to exclude effects of an LSMO electrode. We focus on the electronic states at the Co/PTO interfaces (PTO = PbTiO$_{3}$) and address in particular how the interfacial magnetoelectric coupling affects the local magnetic moments. For this purpose, we compare results for tunnel junctions with one Co/TiO$_{2}$ or one Co/ZrO$_{2}$ interface while keeping a PbO$_{2}$/Co interface.

The Paper is organized as follows. Computational details and the structural models are presented in Sec.~\ref{sec:computational}. In Section~\ref{sec:discussion-results} we discuss our results: structural effects (\ref{sec:structural-effects}), magnetoelectric coupling (\ref{sec:me-coupling}), barrier metalization (\ref{sec:metallization}), and spin-dependent transport (\ref{sec:transport}). We close with a summary and an outlook (Sect.~\ref{sec:summary}).

\section{Structure models and computational details}
\label{sec:computational}
\begin{figure*}
  \centering
  \includegraphics[width = 0.49\textwidth]{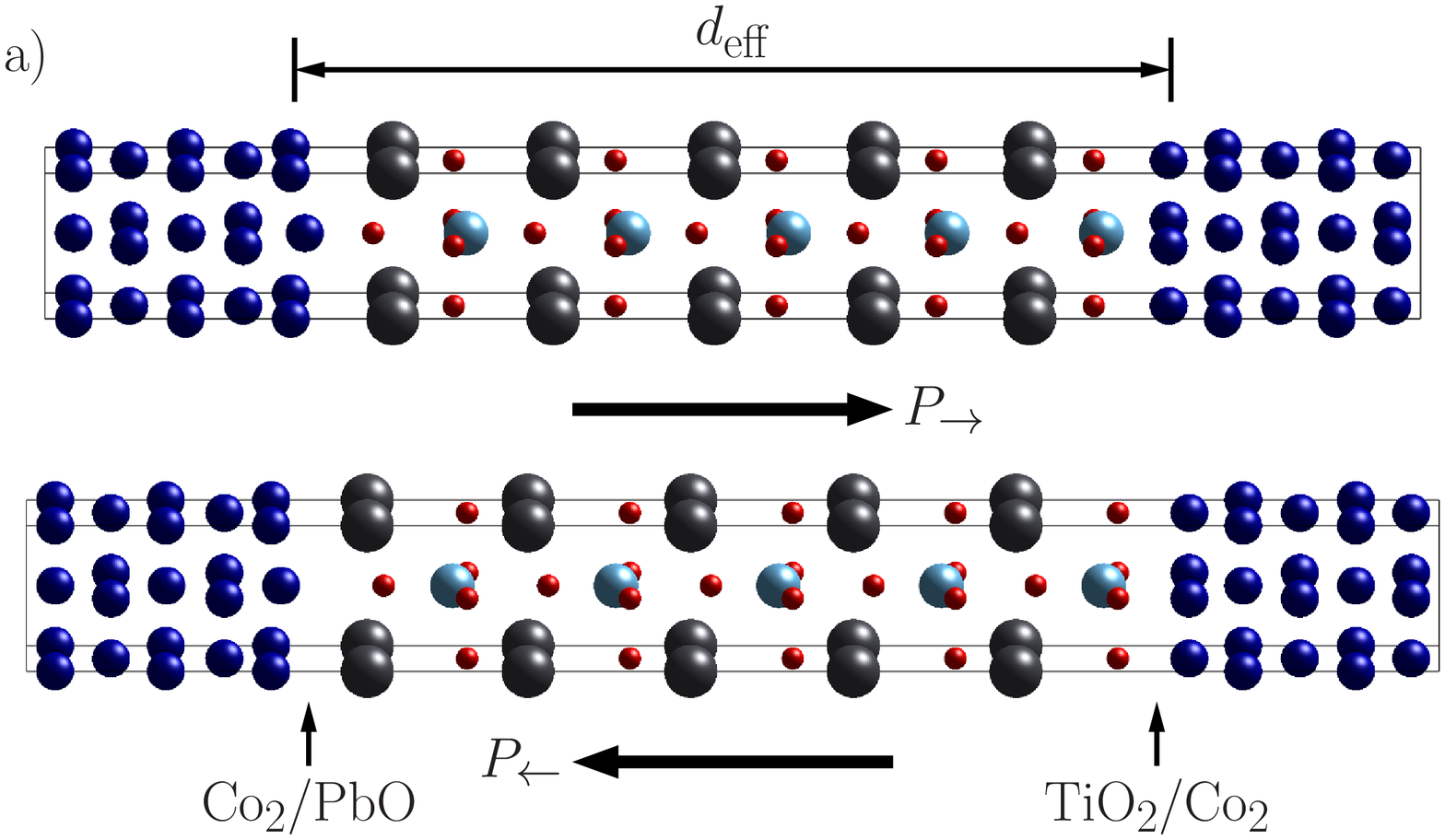}
  \includegraphics[width = 0.49\textwidth]{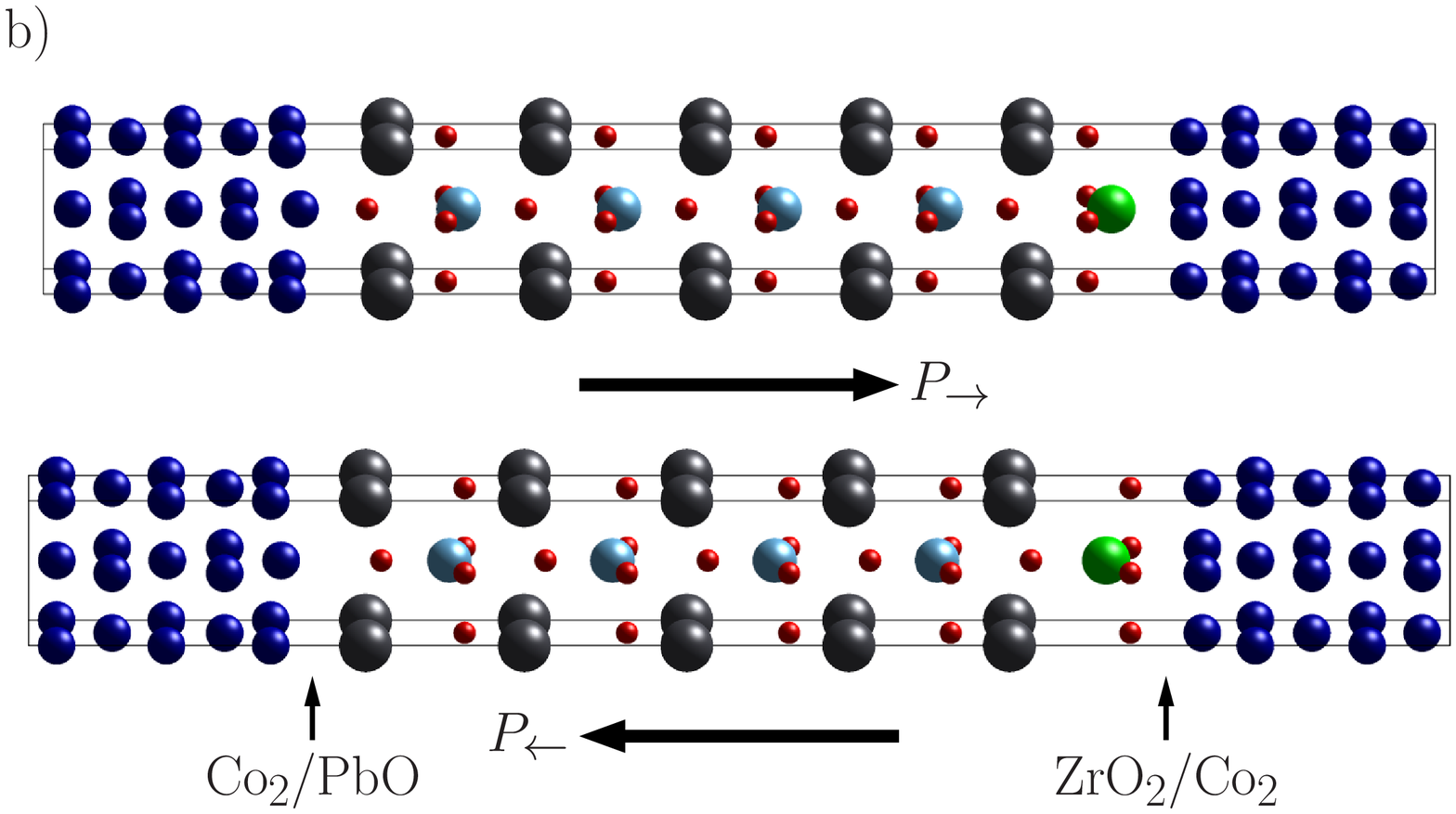}
  \includegraphics[width = 0.8\textwidth]{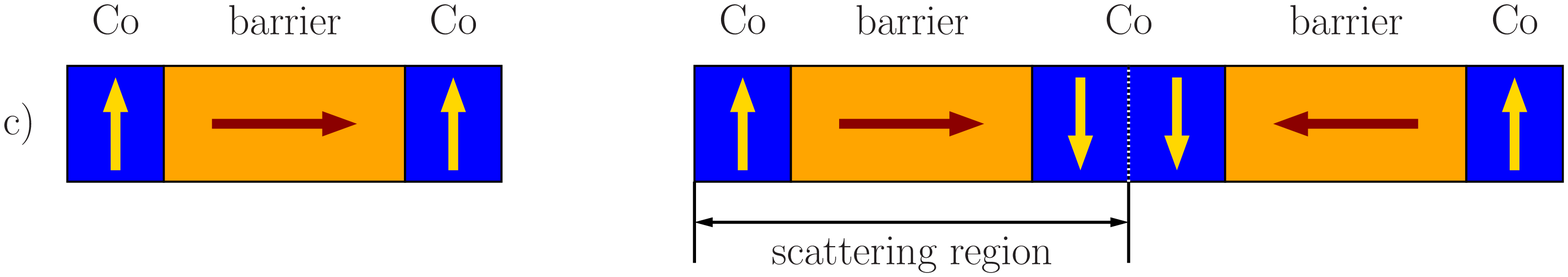}
  \caption{(Color online) Structural models for Co/PTO/Co (a) and Co/PTO-ZO/Co (b) multiferroic tunnel junctions for the two ferroelectric polarizations $P_{\leftarrow}$ and $P_{\rightarrow}$ of the barriers.  The chemical compositions of the interfaces are marked below each structure. The effective thickness  $d_{\mathrm{eff}}$ of the barrier, i.\,e.\ the distance between the Co electrodes, is indicated in (a). In (c), both an original (left, used in the geometry optimization) and a doubled supercell (right, used in the transport calculations) are sketched.}
  \label{f:tj1-2-supercell}
\end{figure*}

The Co/PTO/Co junctions are modeled by supercells (Fig.~\ref{f:tj1-2-supercell}). In analogy to Co films on GaAs substrates \cite{Walmsley1983}, we assume that the Co atoms form a face-centered tetragonal (fct) structure, which is in agreement with epitaxial growth of Co on PTO\@. 

Because the barrier consists of complete PTO unit cells (UCs), in analogy to BaTiO$_{3}$ (Refs.~\onlinecite{Meyerheim2011} and~\onlinecite{Meyerheim2013}), we are concerned with two different interfaces with the Co electrodes on either side of the PTO stack: Co$_{2}$/PbO (left interface) and TiO$_{2}$/Co$_{2}$ (right interface). In agreement with earlier studies of  TiO$_{2}$/Co$_{2}$ interfaces\cite{Oleinik2001}, the most stable configuration has Co placed in line with the interfacial O ions. At a Co$_{2}$/PbO interface, Co atoms are placed in line with Pb and O\@. 

The barrier thickness is five perovskite $AB$O$_{3}$ UCs, the in-plane lattice constant is taken from experiment (PTO: $a = \unit[3.892]{\AA}$, Ref.~\onlinecite{Mestric2005}). At each side of the barrier five atomic layers of fct Co as well as a vacuum region of $\unit[20]{\AA}$ are attached. We define majority and minority spins with respect to the right Co electrode.

To mimic a PZT barrier, we rely on a digital alloy model in which interfacial Ti is replaced by Zr. Thus, Zr forms a chemically complete ZrO$_{2}$ monolayer at the right interface [Fig.~\ref{f:tj1-2-supercell}(b)]. Earlier studies\citep{Borisov2014} showed that a simulation of the $\unit[25]{\%}$ Zr composition in each Ti(Zr)O$_{2}$ layer by in-plane $2\times 2$ supercells gives structural and magnetic properties of the Co/PZT interface which are similar to the digital alloy model used here. In the forthcoming, we refer to this barrier as PTO-ZO\@.

The geometries of the above supercells have been optimized using the \textsc{vasp} code \cite{Kresse1994,Kresse1996}. Initially, the ionic displacements in the ferroelectric were taken as their theoretical bulk values ($\unit[0.45]{\AA}$ and $\unit[0.33]{\AA}$ for PbO and TiO$_{2}$ planes, respectively), in accordance with the ferroelectric polarization being along [001] ($z$ axis).  The fct Co leads have two atoms per layer with an interlayer distance of $\unit[1.47]{\AA}$ along [001] and an in-plane nearest-neighbor distance of $\unit[2.75]{\AA}$. Within the barrier, the atomic positions of the three central UCs were fixed to the bulk values but all atoms near the interfaces were allowed to move along the [001] direction. A $\Gamma$-centered $4 \times 4 \times 2$ Monkhorst-Pack mesh \cite{Monkhorst1976} and an upper limit of $\unit[10^{-2}]{eV / \AA}$ for the ionic forces guarantee accurate structural relaxations. We define $P_{\rightarrow}$ for displacements $\Delta z > 0$ and $P_{\leftarrow}$ for $\Delta z < 0$.

For the transport calculations the above supercells have to be extended. To treat both the parallel (P, $\uparrow\uparrow$) and antiparallel (AP, $\downarrow\uparrow$) alignments of the magnetizations within the Co electrodes, a double supercell was constructed by creating a mirror image of the original supercell with respect to the (001) plane and  attaching it to the original junction [Fig.~\ref{f:tj1-2-supercell}(c)].  These double supercells comprise two PTO barriers whose ferroelectric polarizations are oppositely oriented to each other. Co layers were added to ensure the correct layer alternation. The electronic states in the interior of the Co stacks have to be close to those of Co bulk since later, with respect to an appropriate treatment of electronic transport, two semi-infinite leads will be attached to the scattering region.
 
For the calculation of the tunnel conductances $G$, one half of a double supercell is taken as scattering region. Using the Landauer-B\"{u}ttiker formalism implemented in the \textsc{quantum  espresso} code \cite{Buettiker1985,Smogunov2004,QEpaper2009}, transmission functions $T(k_{x},k_{y})$ have been computed for Co/PTO/Co and Co/PTO-ZO/Co within the two-dimensional Brillouin zone (2BZ). The latter is a square with $-\pi/a \leq k_{x} \leq \pi/a$ and $-\pi/a \leq k_{y}  \leq \pi/a$. An adaptive $\vec{k}$ mesh refinement \cite{Henk2001} reliably  yields accurate tunnel conductances
\begin{align}
  G = \int_{\mathrm{2BZ}} T(k_{x}, k_{y})\,\mathrm{d}k^{2}.
\end{align}
Conductances with an error less than $\unit[10^{-8}]{e^{2} / h}$ are achieved after a few refinements. To resolve all important features of the transmission maps, about $48000$ $\vec{k}$ points in the 2BZ prove sufficient. The smallest triangles in the adaptive integration have an area of about $5 \cdot 10^{-6} (2 \pi/ a)^{2}$.

Having well converged tunnel conductances $G$, the TMR is calculated by
\begin{align}
  \mathrm{TMR}(P) & = \frac{G^{\uparrow\uparrow}_{P} - G^{\downarrow\uparrow}_{P}}{G^{\uparrow\uparrow}_{P} + G^{\downarrow\uparrow}_{P}} \cdot \unit[100]{\%}, \quad P = P_{\leftarrow}, P_{\rightarrow} \label{e:TMR}
\end{align}
for both orientations of the ferroelectric polarization $P$. For each magnetic configuration $M$ of the leads, the TER ratio is given by
\begin{align}
  \mathrm{TER}(M) = \frac{G_{\rightarrow}^{M} - G_{\leftarrow}^{M}}{G_{\rightarrow}^{M} + G_{\leftarrow}^{M}} \cdot \unit[100]{\%}, \quad M = \uparrow\uparrow, \downarrow\uparrow \label{e:TER}
\end{align}
in which  $G_{\rightarrow}^{M}$ ($G_{\leftarrow}^{M}$) corresponds to the conductance for the barrier polarization pointing towards the TiO$_{2}$/Co$_{2}$ (Co$_{2}$/PbO) interface.

Spin-resolved densities of states (DOS) have been calculated for these systems using the \textsc{quantum  espresso} code \cite{QEpaper2009}. We use a $\Gamma$-centered $15\times 15\times 1$ $\vec{k}$ mesh with a smearing of $\unit[0.02]{Ryd}$. Energy cutoffs read $\unit[63]{Ryd}$ for the wavefunctions and $\unit[504]{Ryd}$ for the charge  density. All calculations were performed within the generalized-gradient approximation \cite{Perdew1996} (GGA-PBE) to the exchange-correlation potential.

\section{Discussion and results}
\label{sec:discussion-results}

\subsection{Structural effects}
\label{sec:structural-effects}
The ionic displacements in the PbO and TiO$_{2}$ planes of relaxed PTO are homogeneous along [001] for the $P_{\leftarrow}$ configurations, that is, they do not depend significantly on the layer. For the opposite polarization $P_{\rightarrow}$, these displacements are considerably suppressed near the TiO$_{2}$/Co$_{2}$-terminated interface \cite{Borisov2014}, which is related to the lower stability of this ferroelectric configuration. Zr impurities introduced into the PTO barrier stabilize the local displacements across the entire barrier; the displacement at the `digital' ZrO$_{2}$ layer is sizably increased, in agreement with a previous study of Co/PZT interfaces \cite{Borisov2014}.

The effective thickness $d_{\mathrm{eff}}$ of the ferroelectric barrier is an important property as far as transport properties are concerned. It may simply be defined as the average distance between the terminating Co layers of each electrode [Fig.~\ref{f:tj1-2-supercell}(a)]. We find that $d_{\mathrm{eff}}$ is reduced for $P_{\rightarrow}$ in comparison to $P_{\leftarrow}$: $\unit[1.1]{\AA}$ for PTO and $\unit[0.8]{\AA}$ for PTO-ZO\@. This observation is in agreement with recent transport measurements of Co/PZT/LSMO tunnel junctions \cite{Quindeau}. It is explained by the bonding at the interfaces. Whereas the Co-O and Ti(Zr)-Co bonds dominate at the \textit{B}O$_{2}$/Co$_{2}$ side (\textit{B} = Ti, Zr), the Pb-Co and Co-O chemical bonds determine the geometry of the Co$_{2}$/PbO interface. In both cases, the ferroelectric switching leads to either contraction or expansion of these bonds, thereby changing $d_{\mathrm{eff}}$.

\subsection{Magnetoelectric coupling}
\label{sec:me-coupling}
\begin{figure}
  \centering
  \includegraphics[width = \columnwidth]{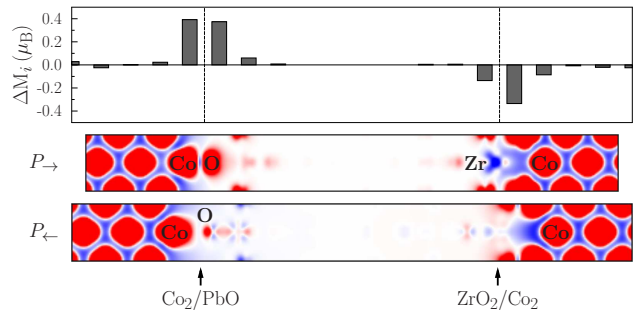}
  \caption{(Color online) Magnetoelectric coupling in a Co/PTO-ZO/Co tunnel junction. Top: difference of the layer-resolved magnetic moments $M_{i}$ upon reversal of the ferroelectric polarization in the barrier ($P_{\leftarrow}$ and $P_{\rightarrow}$): $\Delta M_{i} \equiv M_{i}(P_{\rightarrow}) - M_{i}(P_{\leftarrow})$, $i$ layer index. Bottom: spin-resolved charge densities for the two ferroelectric polarizations (majority: red; minority: blue).}
  \label{f:tj2-spin-density}
\end{figure}

The geometrical changes that accompany the reversal of the ferroelectric polarization affect the magnetic structure at the interfaces as well. For a Co/PTO-ZO/Co MFTJ we find sizable magnetic moments induced on the O ions on the left side and on the Zr cations on the right side in comparison with those of the Co leads ($\unit[1.8]{\mu_{\mathrm{B}}}$; Fig.~\ref{f:tj2-spin-density}). This behavior is present also in Co/PTO/Co junctions, for which even larger magnetic moments are induced on the Ti cations at the TiO$_{2}$/Co$_{2}$ interface. These findings indicate hybridization of orbitals at the interfaces, in particular those involved in the Ti-Co and Co-O bonds \cite{Borisov2014}. As a result, a strong magnetoelectric (ME) coupling is found for both interfaces.

A similar mechanism for the ME coupling is expected for the Co$_{2}$/PbO interface. However, in this case, the effect is attributed mostly to the Co-O bonds. As the polarization is switched from $P_{\leftarrow}$ to $P_{\rightarrow}$, the Co-O bonds are shortened by more than $\unit[30]{\%}$ (from $\unit[2.65]{\AA}$ to $\unit[1.78]{\AA}$). The Co-Pb bonds, on the other hand, are moderately stretched from $\unit[2.51]{\AA}$ to $\unit[2.65]{\AA}$. The combined effect of these structural changes is a buckling of the interfacial Co layers. Accordingly, the magnetic moments of the two Co species differ: the magnetic moments of Co atoms in line with Pb cations are considerably decreased by about $\unit[-0.3]{\mu_{\mathrm{B}}}$ for $P_{\leftarrow}$; simultaneously, those of the neighboring Pb cations are increased by about $\unit[0.2]{\mu_{\mathrm{B}}}$. The Co magnetic moments close to the interface vary between $\unit[1.5]{\mu_{\mathrm{B}}}$ and $\unit[1.9]{\mu_{\mathrm{B}}}$, depending on layer and ferroelectric polarization.

\begin{figure}
  \centering
  \includegraphics[width = \columnwidth]{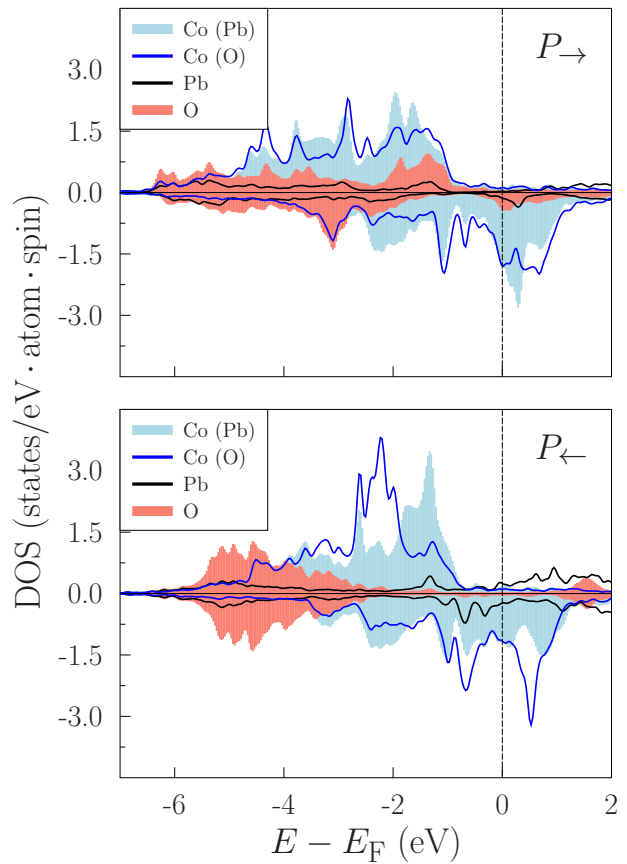}
  \caption{(Color online) Effect of the ferroelectric polarization $P$ in a Co/PZT/Co tunnel junction on the electronic structure at the interfaces for $P_{\rightarrow}$ (top) and $P_{\leftarrow}$ (bottom) configurations. Spin-resolved densities of states are depicted for Co atoms in line with Pb [Co (Pb)] and in line with O [Co (O)], as well as for O and Pb cations at the Co$_{2}$/PbO interface.}
  \label{f:DOS-IF1-comparison}
\end{figure}

The above effect is due to an increased overlap between the electronic states of these cations; confer the site-projected density of states in Fig.~\ref{f:DOS-IF1-comparison}. The magnetic moment of Co in line with O ions is less affected. However, O-p orbitals hybridize with Co-d orbitals in a broad energy range. Furthermore, the charge density becomes more delocalized on the O sites for $P_{\rightarrow}$ (top in Fig.~\ref{f:DOS-IF1-comparison}). The larger orbital overlap of the aforementioned states enhances the induced moments on the O sites which are parallel to those of the neighboring Co atoms. Therefore, the ME coupling at the Co$_{2}$/PbO interface relies on the induced magnetic moments in the PbO layer and on the change of the Co magnetic moments. This strong localization of the ME coupling to the interface is evident from the layer-resolved magnetization in the Co/PTO-ZO/Co heterostructure (Fig.~\ref{f:tj2-spin-density}).

The strength of the ME coupling is quantified by the magnetoelectric coupling constant $\alpha = \Delta M / (A E_{\mathrm{c}})$. It is defined as the change  $\Delta M$ of the magnetization in the interface area $A$ that is induced by the coercive electric field $E_{\mathrm{c}}$ (that is the minimum field strength needed to switch the ferroelectric polarization). From the data presented in Fig.~\ref{f:tj2-spin-density}, we estimate $\alpha$ to $\unit[9.1 \cdot 10^{-10}]{G\,cm^{2} V^{-1}}$ for the Co$_{2}$/PbO and to $\unit[-5.7 \cdot 10^{-10}]{G\,cm^{2} V^{-1}}$ for the TiO$_{2}$/Co$_{2}$ interface, assuming the coercive field of bulk PTO ($E_{\mathrm{c}} = \unit[33]{kV / cm}$; the coercive field for the PTO-ZO barrier might be smaller). These numbers are in line with those reported for similar heterostructures\cite{Velev2011} (e.\,g.\ Co/PZT, Ref.~\onlinecite{Borisov2014}; LSMO/PTO, Ref.~\onlinecite{Borisov2015a}; Fe/BaTiO$_{3}$, Ref.~\onlinecite{Duan2006}).

Summarizing at this point, a strong ME coupling of electronic origin is established for both interfaces in the undoped and in the Zr-doped Co/PTO/Co junctions. The two interfaces differ with respect to the sign of the magnetization change in response to the ferroelectric switching. We find no  qualitative change upon Zr doping of the PTO barrier with respect to the undoped barrier. The ME coupling is traced back to the polarization-dependent hybridization of orbitals that are involved in the Co-O, Pb-Co, and Ti(Zr)-O bonds. Because the detailed electronic structure at the interfaces determines significantly the spin-dependent transport, we expect a considerable effect on the electron transmission across the junctions.

\subsection{Barrier metalization}
\label{sec:metallization}
Before discussing spin-dependent transport, we address whether the barriers are insulating (tunneling regime) or conducting (metallic regime). For this purpose, we investigate the site-resolved density of states at the Fermi level; a layer is considered insulating if this quantity is negligibly small.

\begin{figure*}
  \centering
  \includegraphics[width = 0.9\textwidth]{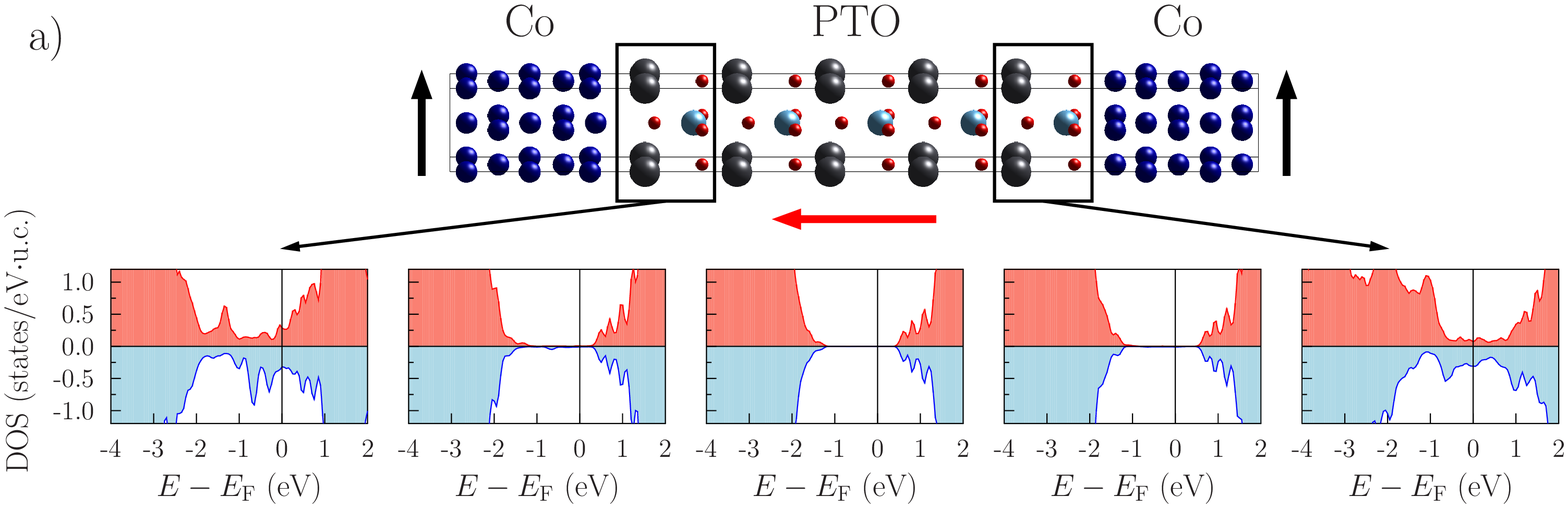}
  \includegraphics[width = 0.9\textwidth]{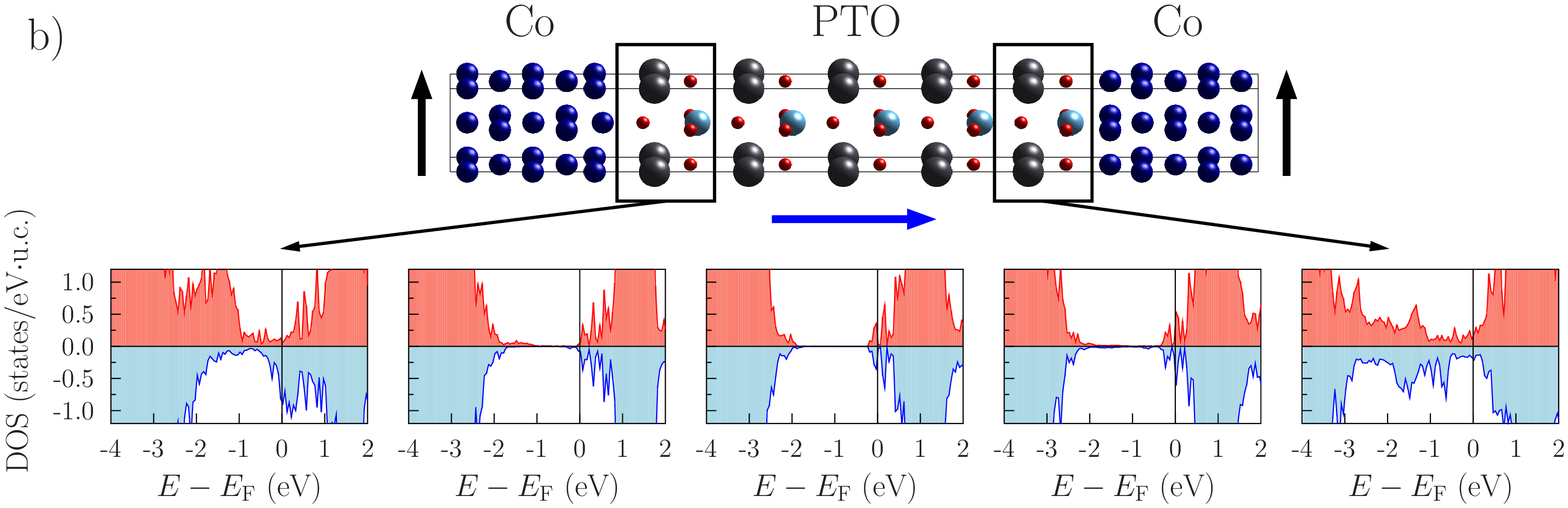}
  \includegraphics[width = 0.9\textwidth]{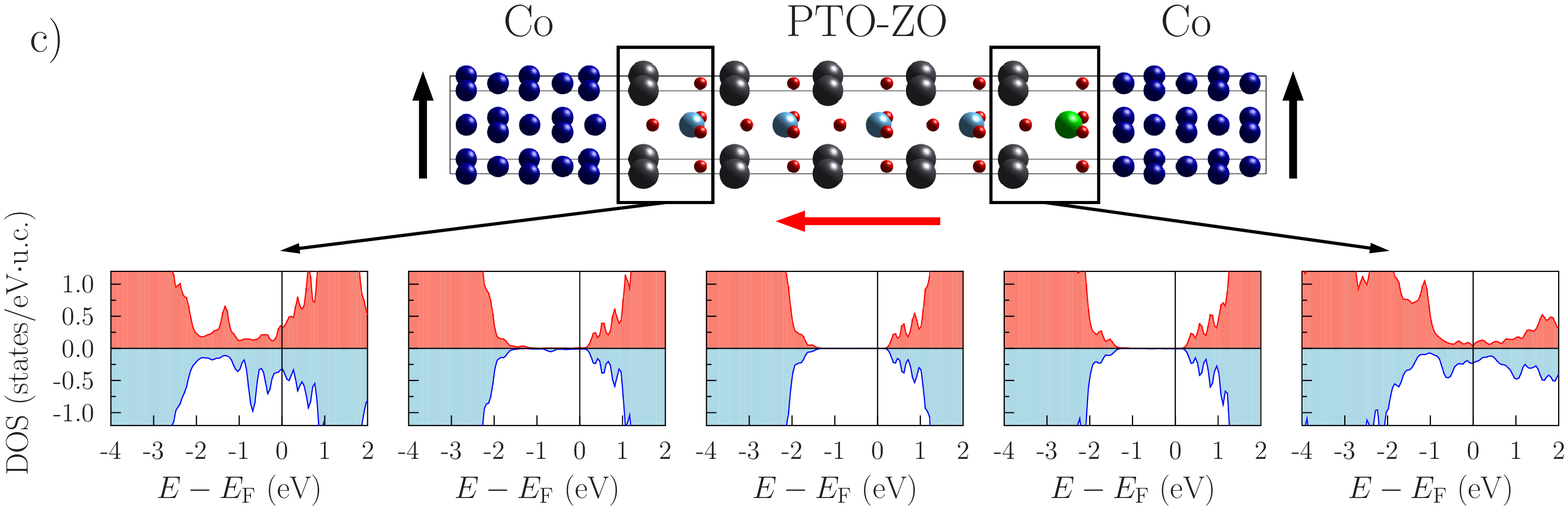}
  \includegraphics[width = 0.9\textwidth]{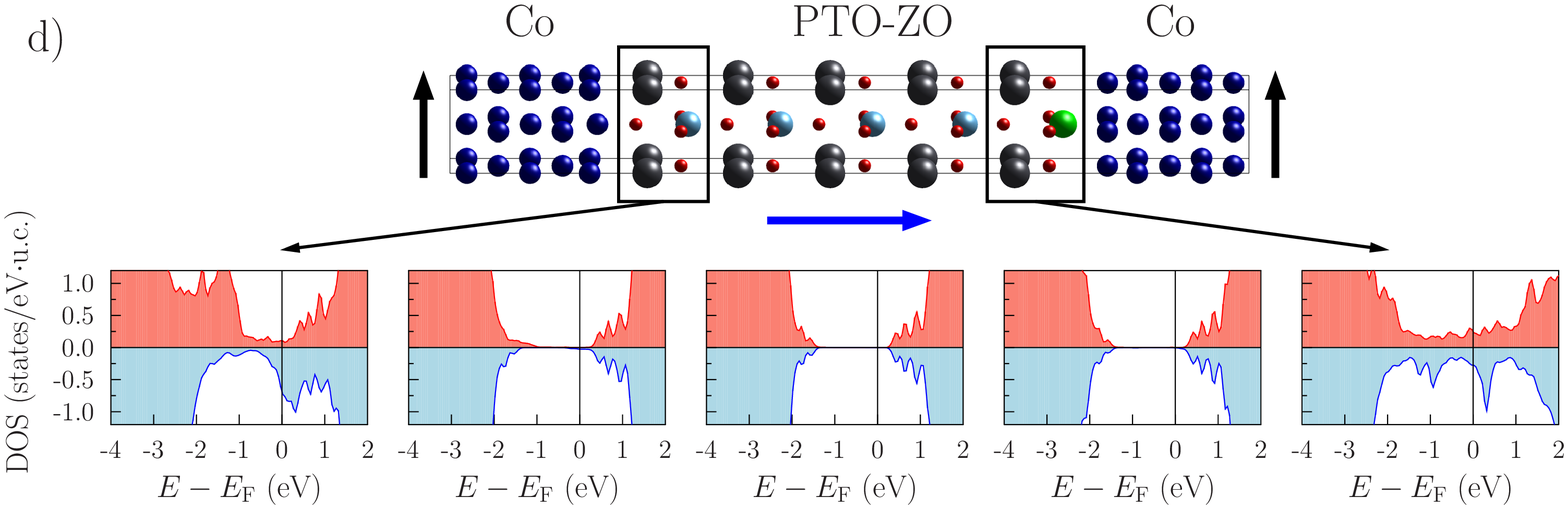}
  \caption{(Color online) Electronic structure of Co/PTO/Co [upper two panels, (a) and (b)] and Co/PTO-ZO/Co [lower two panels, (c) and (d)] tunnel junctions for parallel alignment of the lead magnetizations (black vertical arrows, $\uparrow\uparrow$) and for the two polarizations $P_{\leftarrow}$ [(a) and (c)] and $P_{\rightarrow}$ [(b) and (d)] in the ferroelectric barrier (as indicated in each panel by horizontal arrows). Spin-resolved densities of states are shown for the series of unit cells across the barriers; majority spin red, minority spin blue.}
  \label{f:tj1-2-barrier-DOS}
\end{figure*}

For the $P_{\leftarrow}$ configuration, the PTO barrier is locally metallic within one UC at the Co$_{2}$/PbO interface and in the TiO$_{2}$ plane at the right interface; the inner 3.5 UCs are insulating [Fig.~\ref{f:tj1-2-barrier-DOS}(a)]. This partial metalization is due to charge transfer from Co atoms into the ferroelectric, which happens at both interfaces. For $P_{\rightarrow}$, the effect is even enhanced, so that the entire barrier becomes metallic [Fig.~\ref{f:tj1-2-barrier-DOS}(b)] and one expects very large conductances (as compared to those for the tunneling regime).

The metalization affects mostly the TiO$_{2}$ layers but leaves the PbO layers insulating. An exception is the terminating PbO layer at the left interface, in which sizable magnetic moments are induced on the O sites. Such an alternating sequence of conducting TiO$_{2}$ and insulating PbO layers might open new possibilities for electron transport.

In contrast to the cases discussed above, the Zr-doped barriers are insulating irrespectively of the ferroelectric polarization [Figs.~\ref{f:tj1-2-barrier-DOS}(c) and (d)] and, thus, we are concerned with the tunneling regime. More precisely, metalization is restricted to a single UC at each interface, similarly to Co/PTO/Co for $P_{\leftarrow}$ [Fig.~\ref{f:tj1-2-barrier-DOS}(a)]. It is conceivable that this fundamental difference of the PTO- and PTO-ZO-based junctions is correlated with the aforementioned stabilization of the ionic displacements in the barrier by Zr impurities.

\subsection{Spin-dependent transport}
\label{sec:transport}
We now discuss spin-dependent transport for the two types of tunnel junctions in four different configurations.

\subsubsection{Co/PTO/Co}
\begin{figure*}
  \centering
  \includegraphics[width = 0.45\textwidth]{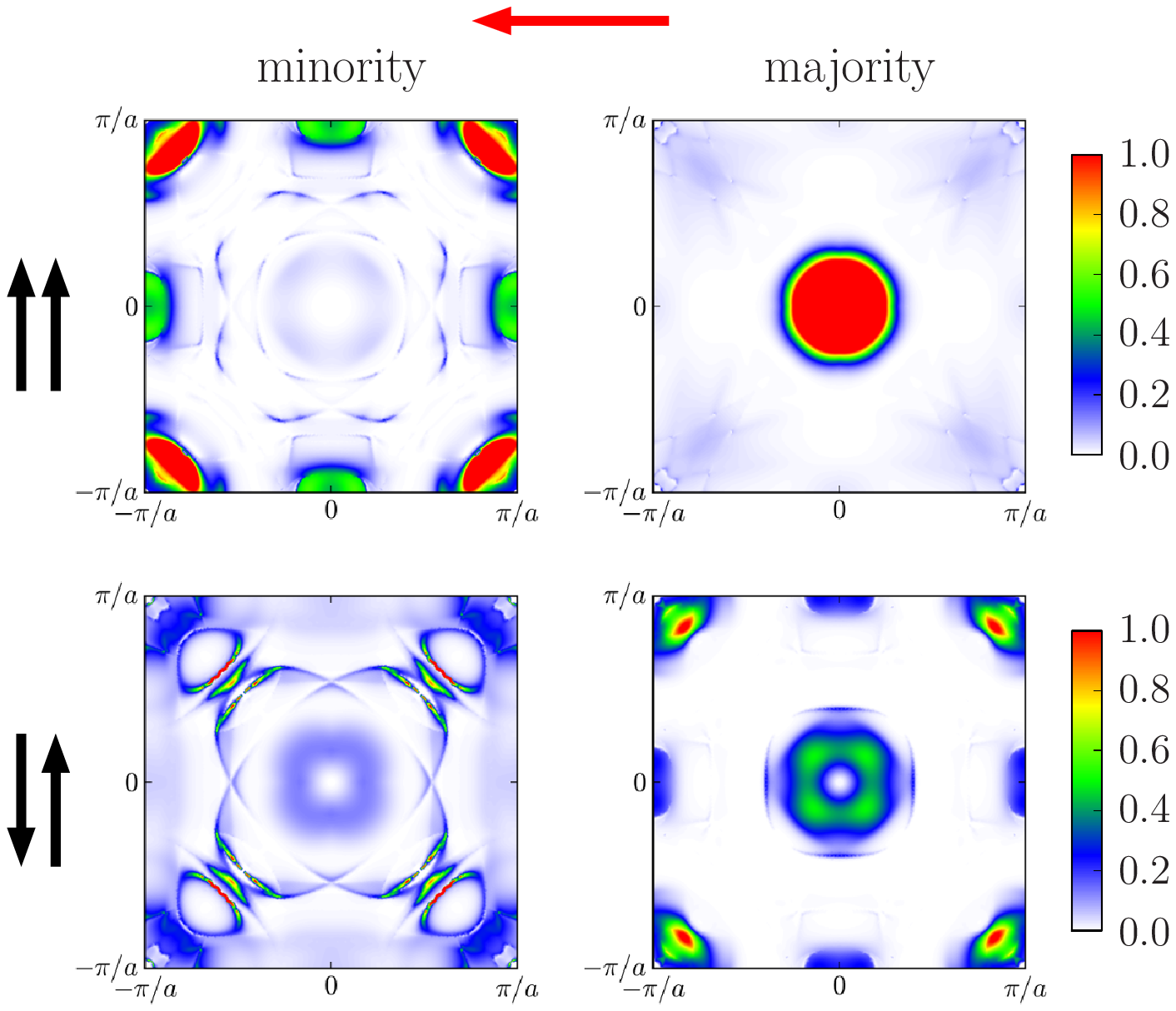}
  \includegraphics[width = 0.45\textwidth]{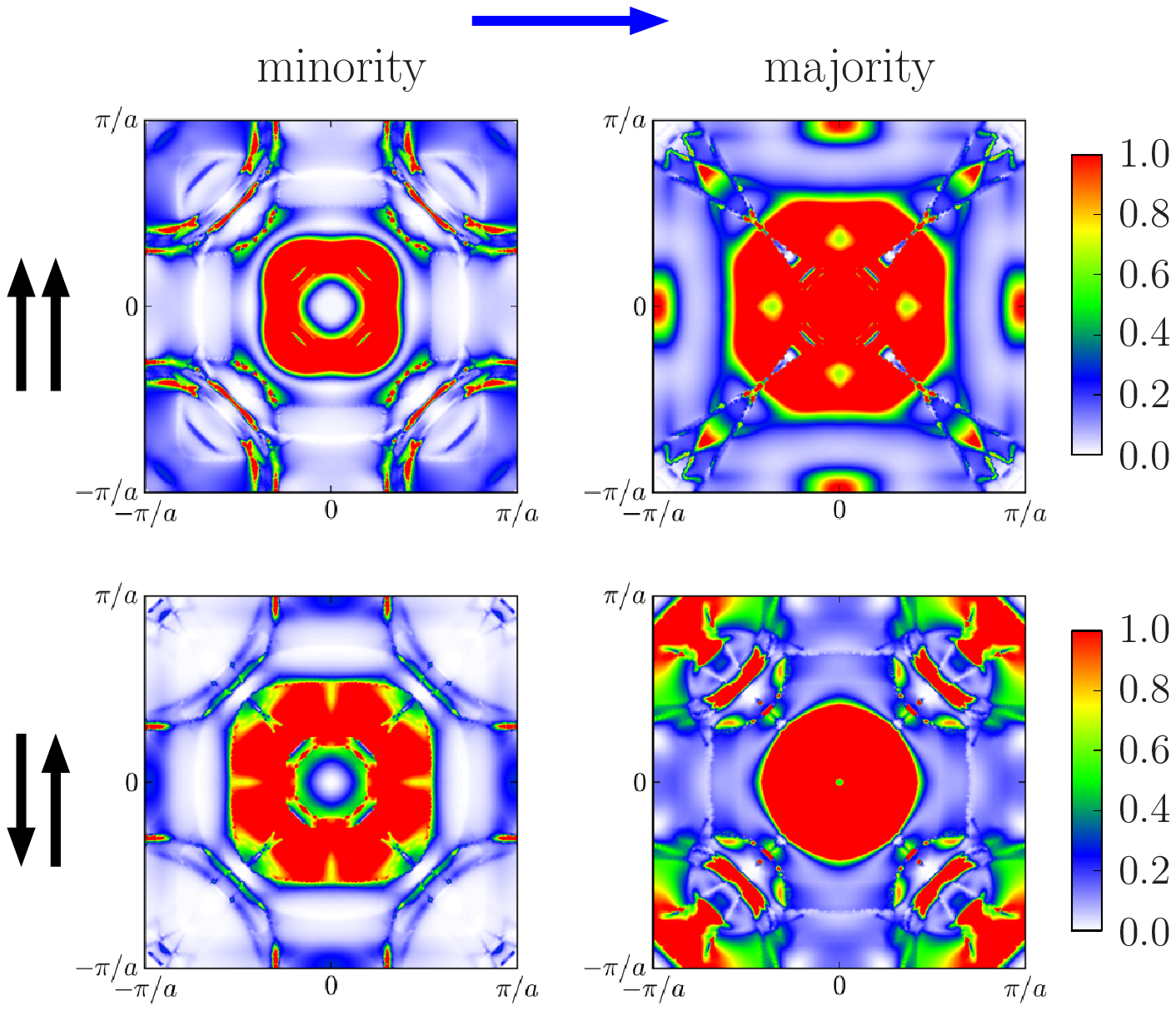}\\
  \includegraphics[width = 0.45\textwidth]{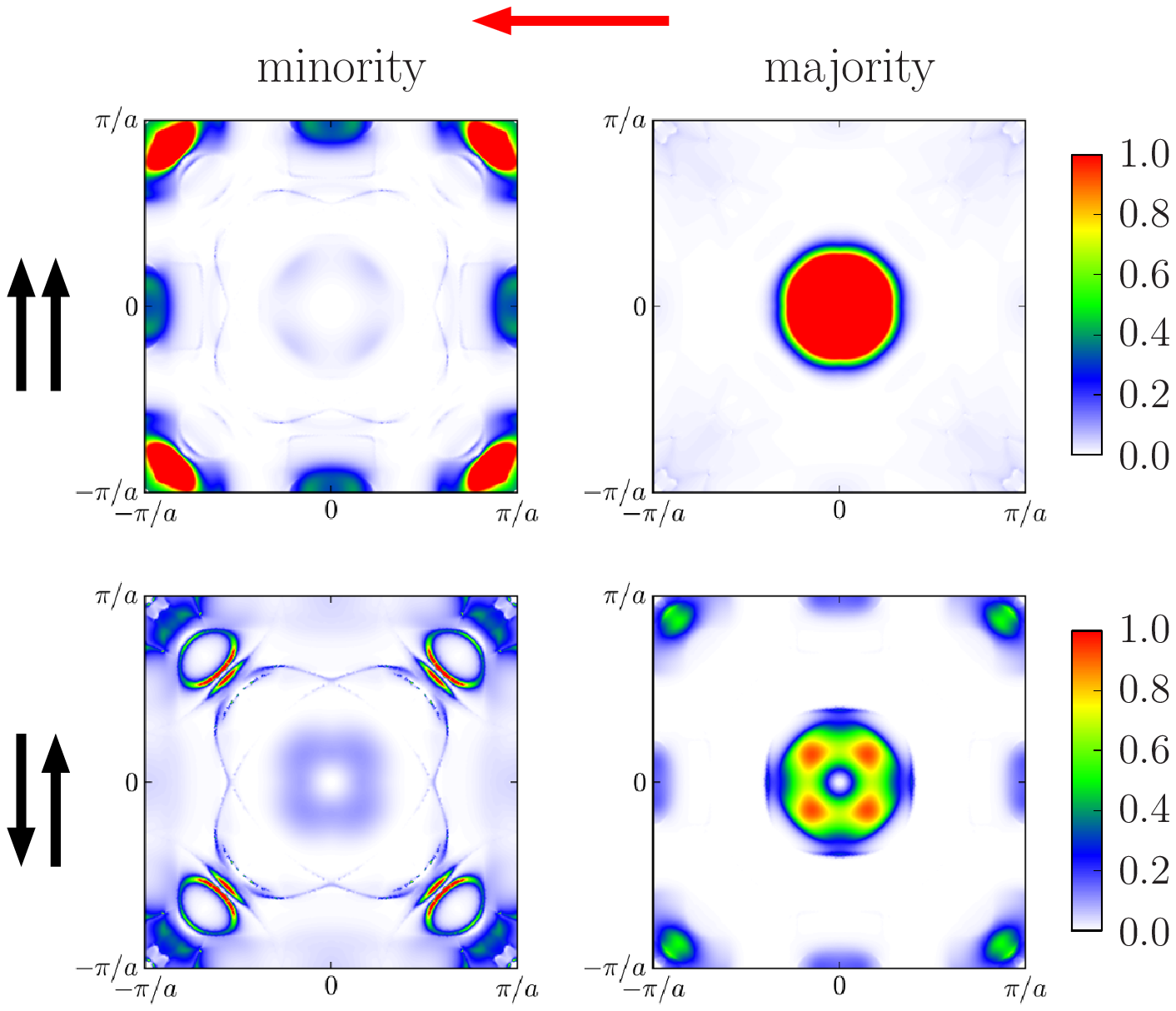}
  \includegraphics[width = 0.45\textwidth]{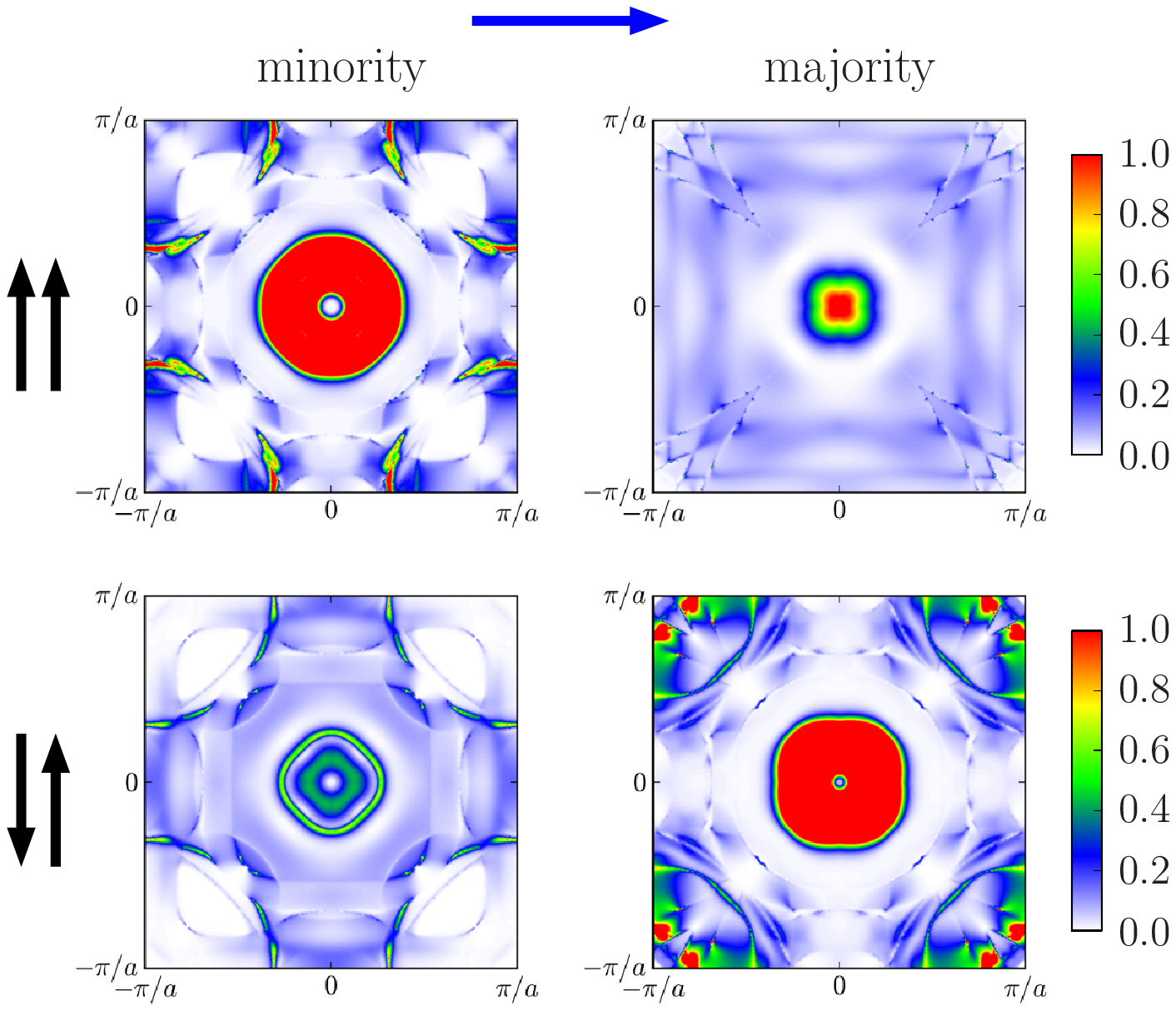}
  \caption{(Color online) Spin-dependent transport in Co/PTO/Co (top row) and Co/PTO-ZO/Co (bottom row) tunnel junctions. In each panel, transmission maps $T(k_{x}, k_{y})$ are shown as a color scale for the parallel ($\uparrow\uparrow$) and the antiparallel ($\downarrow\uparrow$) configuration of magnetic Co electrodes as well as for the two ferroelectric polarizations of the barriers (indicated by horizontal arrows). Color scales in units of $\unit[10^{-5}]{e^{2} / h}$.}
  \label{f:tj1-2-transmission}
\end{figure*}

For the Co/PTO/Co tunnel junction, the transmittances $T(k_{x}, k_{y})$ reveal a drastic change of the transmission (upper part of Fig.~\ref{f:tj1-2-transmission}): for $P_{\leftarrow}$ the ferroelectric PTO barrier is insulating, whereas for $P_{\rightarrow}$ it is conducting, as shown in Fig.~\ref{f:tj1-2-barrier-DOS}. In the latter case, large parts of the 2BZ show very high transmittances (red regions).

The transmittance maps for $P_{\leftarrow}$ (upper left part of Fig.~\ref{f:tj1-2-transmission}) manifest that the parallel magnetic configuration provides larger conductances than the antiparallel configuration, as is also evident from the numbers given in Table~\ref{t:total-transmission-PTO}. Thus, we are concerned with a normal rather than with an inverted TMR\@.

\begin{table}
 \caption{Spin-resolved conductances $G$ of Co/PTO/Co multiferroic tunnel junctions (in units of $\unit[10^{-6}]{e^{2} / h}$) for the parallel ($\uparrow\uparrow$) and the antiparallel ($\downarrow\uparrow$) magnetic configurations of the Co electrodes as well as for the two ferroelectric polarization of the barriers ($P_{\rightarrow}$ and $P_{\leftarrow})$. Minority (min.) and majority (maj.) spin channels are defined with respect to the right Co electrode. The respective tunnel magnetoresistances (TMR) and tunnel electroresistances (TER) are given in addition.}
   \centering
   \setlength{\tabcolsep}{5pt}
   \renewcommand{\arraystretch}{1.2}
   \begin{tabular}{c|cc|cc|c}\hline\hline
     {PTO} & \multicolumn{2}{c|}{$P_{\leftarrow}$} & \multicolumn{2}{c|}{$P_{\rightarrow}$} & {} \\ 
     {} & $G_{\mathrm{min}}$ & $G_{\mathrm{maj}}$ & $G_{\mathrm{min}}$ & $G_{\mathrm{maj}}$ & TER \\ \hline
      $\uparrow\uparrow$ & 1.40 & 2.03 & {159} & {11.3} & \: 96.1\%  \\ \hline
      $\downarrow\uparrow$ & 0.82 & 0.80 & {15.7} & {117.8} & \: 97.6\%  \\ \hline
     {TMR} & \multicolumn{2}{c|}{36\%} & \multicolumn{2}{c|}{12\%} & {} \\ \hline\hline
   \end{tabular}
   \label{t:total-transmission-PTO}
 \end{table}

For the parallel configuration of the electrodes ($\uparrow\uparrow$), transport is dominated by free-electron-like majority electrons. The largest transmission is observed near the center of the 2BZ\@. The minority channel also contributes to the transmission but shows a complicated behavior, with the largest transmission at the $\overline{\mathrm{X}}$ [$\vec{k}(\overline{\mathrm{X}}) = (\nicefrac{\pi}{a}, 0)$ and equivalent] and at the $\overline{\mathrm{M}}$ [$\vec{k}(\overline{\mathrm{M}}) = (\nicefrac{\pi}{a}, \nicefrac{\pi}{a})$ and equivalent] points of the 2BZ\@. The antiparallel configuration ($\downarrow\uparrow$) is characterized by reduced transmittances. Here, both spin channels contribute almost equally to the conductance.

Turning to the $P_{\rightarrow}$ configuration (upper right part of Fig.~\ref{f:tj1-2-transmission}), the transmission maps change dramatically; we recall that here the barrier becomes conducting, leading to sizable areas in the 2BZ with high transmittance. There are also large contributions that arise from resonant tunneling. As a result, the conductances are increased by two orders of magnitude as  compared to the tunneling regime, yielding giant TER ratios (almost $\unit[100]{\%}$, Table~\ref{t:total-transmission-PTO}).

It turns out that also for the metallic PTO barrier the TMR is not inverted, in contrast to experiment. A TMR ratio of $\unit[12]{\%}$ is significantly lower than for $P_{\leftarrow}$ ($\unit[36]{\%}$).

We conclude that---although being illustrative with respect to effects of insulating and conducting barriers---the PTO barrier cannot model reliably the experimental observations.

\subsubsection{Co/PTO-ZO/Co}
The effect of Zr doping on the electronic transport is analyzed by comparing the transmittances of Co/PTO/Co with those of the Co/PTO-ZO/Co digital alloy.

For $P_{\leftarrow}$ (bottom left part of Fig.~\ref{f:tj1-2-transmission}), the transmission for the parallel configuration is enhanced in the majority spin channel for the Zr-doped system, whereas it is mildly affected in the minority spin channel (cf.\ Table~\ref{t:total-transmission-PTO-ZO}). This observation is at variance with an admittedly very simple explanation by the effective thickness of the ferroelectric barrier; the latter increases upon Zr doping, which would usually lead to lower transmittances because of the exponential decay of electronic states within the barrier. One could argue that the larger extent of the Zr-d orbitals compared to the Ti-d orbitals might increase the transmission probability across the ZrO$_{2}$/Co$_{2}$ interface due to the stronger hybridization with the orbitals at neighboring atomic sites.

\begin{table}
 \caption{As Table~\ref{t:total-transmission-PTO} but for Co/PTO-ZO/Co multiferroic tunnel junctions.}
   \centering
   \setlength{\tabcolsep}{5pt}
   \renewcommand{\arraystretch}{1.2}
   \begin{tabular}{c|cc|cc|c}\hline\hline
     {PTO-ZO} & \multicolumn{2}{c|}{$P_{\leftarrow}$} & \multicolumn{2}{c|}{$P_{\rightarrow}$} & {} \\ 
      {} & $G_{\mathrm{min}}$ & $G_{\mathrm{maj}}$ & $G_{\mathrm{min}}$ & $G_{\mathrm{maj}}$ & TER \\ \hline
      $\uparrow\uparrow$ & 1.43 & 3.55 & 16.2 & 0.79 & \: 55\% \\ \hline
      $\downarrow\uparrow$ & 0.70 & 0.82 & 0.80 & 8.07 & \: 71\% \\ \hline
      {TMR} & \multicolumn{2}{c|}{53\%} & \multicolumn{2}{c|}{31\%} & {} \\ \hline\hline
   \end{tabular}
   \label{t:total-transmission-PTO-ZO}
 \end{table}

A significant change of the conductances upon Zr doping is not observed for the antiparallel lead magnetizations (Table~\ref{t:total-transmission-PTO-ZO}). In this case, the Zr impurities result in a redistribution of the transmittances but keep the general shape of the $T$ maps. 

The transmittance maps for $P_{\rightarrow}$ display an eye-catching difference to those for $P_{\leftarrow}$. Nevertheless, the sign of the TMR is preserved and its absolute value is moderately reduced from $\unit[53]{\%}$ to $\unit[31]{\%}$. We recall that for $P_{\rightarrow}$ we are concerned with tunneling, in contrast to the Co/PTO/Co junction.

At the Fermi level, the electronic properties of Co are mainly determined by the majority-spin s and the minority-spin d states. Although the minority states dominate in bulk Co at the Fermi level, our \textit{ab initio} calculations for $P_{\leftarrow}$ tunnel junctions suggest that these electrons are transmitted less than the majority electrons.

Concerning $P_{\rightarrow}$, the electron (hole) charge transfer across the interface plays a crucial role. We found that a  large (in absolute value) negative spin polarization is induced on the interfacial Ti/Zr-d orbitals, which originates from the hybridization with the neighboring d orbitals of Co (cf.\ Figs.~11 and~12 in Ref.~\onlinecite{Borisov2014}). As a result, the by far largest conductances show up in the minority channel for $\uparrow\uparrow$ and in the majority channel for $\downarrow\uparrow$ (Table~\ref{t:total-transmission-PTO-ZO}).

The TER is explained by the distribution of free charge carriers in this heterostructure. According to the layer-resolved DOS (Fig.~\ref{f:tj1-2-barrier-DOS}), an appropriately oriented ferroelectric polarization of the barrier induces charge transfer at the TiO$_{2}$/Co$_{2}$ interface which can lead to enhanced propagation of electronic states. This effect would reduce the decay length of electrons within the barrier for $P_{\rightarrow}$ as compared to $P_{\leftarrow}$. Since the tunneling probability  increases exponentially with decrease of the barrier thickness, the above-mentioned partial metalization leads to larger total conductances for $P_{\rightarrow}$. This scenario is confirmed  by our \textit{ab initio} results (Fig.~\ref{f:tj1-2-transmission} and Table~\ref{t:total-transmission-PTO-ZO}).

\section{Summary}
\label{sec:summary}
The present \textit{ab initio} study clarifies the origin of the tunnel magnetoresistance (TMR) and the tunnel electroresistance (TER) in multiferroic Co/PbTiO$_{3}$/Co. The results may be `transferred' to similar heterostructures, i.\,e.\ those that combine ferromagnetic electrodes (e.\,g.\ Fe, Co, and Ni) and well-established ferroelectrics (e.\,g.\ BaTiO$_{3}$). The strong magnetoelectric coupling at the Co/PbTiO$_{3}$ interfaces is explained by spin-dependent hybridization of orbitals; it affects considerably the local magnetic moments and the spin-resolved charge density in a narrow region about the interface. As a result, we find four distinctly different conductances, that is a four-state memory device. 
 
The charge transfer from Co into the ferroelectric barrier results in a metalization of the latter. This transfer is particularly strong for the ferroelectric polarization pointing towards the TiO$_{2}$/Co$_{2}$ interface. Such a complete metalization may be attributed to the local-density approximation in the underlying density-functional calculations which show up as a too small  fundamental band gap ($\unit[1.7]{eV}$ in the bulk but about $\unit[1.5]{eV}$ within the barrier; experiment: $\unit[3.4]{eV}$, from Ref.~\onlinecite{Piskunov04} and references therein): if the calculated band gap would approach the experimental gap \cite{Borisov2015a}, the insulating region of the PbTiO$_{3}$ barrier would be almost as wide as its nominal chemical thickness. This issue needs to be clarified in a future investigation.
 
The inverted TMR observed in experiments on Co/PbZr$_{0.2}$Ti$_{0.8}$O$_{3}$/LaSrMnO$_{3}$ tunnel junctions \cite{Pantel2012}
is not reproduced by calculations for the digitally Zr-doped PbTiO$_{3}$ barrier sandwiched between Co electrodes. This finding suggests that La(Sr)MnO$_{3}$ plays a crucial role for the TMR inversion. Within this respect, the almost half-metallic MnO$_{2}$/PbO interface, instead of Co$_{2}$/PbO considered in this study, may significantly modify the selection of states for spin-dependent transmission (`symmetry filtering') \cite{Burton2009,Burton2012,Chen2012}. Regarding the PbZr$_{0.2}$Ti$_{0.8}$O$_{3}$ barrier, randomly distributed Zr dopants may influence the tunneling as well, since Zr enhances locally the ferroelectric displacements in PbTiO$_{3}$ and modifies the bonding at the interfaces \cite{Borisov2014}. These issues will treated in a future theoretical study.

\acknowledgments
This work is supported by SFB 762 of DFG\@. Parts of some figures have been produced with \textsc{vesta3} (Ref.~\onlinecite{Momma11}).

\bibliographystyle{apsrev}
\bibliography{./Co-PZT-Co}

\end{document}